\documentclass[twocolumn,showpacs,prb]{revtex4}
\usepackage{bm}
\usepackage{graphicx}

\usepackage{amsmath}
\usepackage{amsfonts}
\newcommand{\nix}[1]{}

\begin{document}

\title{Spin-dependent electron dynamics and recombination in GaAs$_{1-x}$N$_{x}$ alloys at room temperature}
\author{V.K.~Kalevich}
\author{A.Yu.~Shiryaev}
\author{E.L.~Ivchenko}
\author{A.Yu.~Egorov}
\affiliation{A.F.~Ioffe Physico-Technical Institute, St.
Petersburg 194021, Russia}
\author{L.~Lombez}
\author{D.~Lagarde}
\author{X.~Marie}
\author{T.~Amand}
\affiliation{LNMO-INSA, 135 avenue de Rangueil, 31077 Toulouse,
France}

\begin{abstract}
We report on both experimental and theoretical study of
conduction-electron spin polarization dynamics achieved by pulsed
optical pumping at room temperature in GaAs$_{1-x}$N$_{x}$ alloys
with a small nitrogen content ($x$ = 2.1, 2.7, 3.4\%). It is found
that the photoluminescence circular polarization determined by the
mean spin of free electrons reaches 40-45\% and this giant value
persists within 2 ns. Simultaneously, the total free-electron spin
decays rapidly with the characteristic time $\approx 150$ ps. The
results are explained by spin-dependent capture of free conduction
electrons on deep paramagnetic centers resulting in dynamical
polarization of bound electrons. We have developed a nonlinear
theory of spin dynamics in the coupled system of spin-polarized
free and localized carriers which describes the experimental
dependencies, in particular, electron spin quantum beats observed
in a transverse magnetic field.
\end{abstract}


\maketitle

\section{INTRODUCTION} Spintronics devices require generation of
high electron spin polarization in non-resonant excitation
conditions and its conservation that must be sufficient to store
and manipulate spin information. We show here that these
requirements can be achieved at room temperature in dilute nitride
III-V semiconductor heterostructures under optical pumping. For
this, we use spin-dependent capture of free electrons by deep
paramagnetic centers present in semiconductor crystal. The
spin-dependent capture results in dynamical polarization of bound
electrons which, in its turn, acts as a spin filter increasing
polarization of free
electrons~\cite{Weisbuch,Miller,Paget,JETPLett}. We demonstrate
here, that, after a short optical pump pulse, the decay of both a
difference of the free-electron spin-up and spin-down densities
and their sum is controlled by the same fast spin relaxation in
the conduction band. As a result, the spin polarization of free
electrons is constant while their total spin decays fast when the
time delay increases.

\section{Experimental results}   The samples under study are
undoped 0.1-$\mu$m-thick GaAs$_{1-x}$N$_{x}$ ($x$ = 2.1, 2.7,
3.5\%) layers grown by MBE on a (001) GaAs
substrate~\cite{Egorov}. We investigate electron spin properties
by polarized time-resolved photoluminescence (PL). The samples are
excited along the growth axis ($z$-axis) by circularly
($\sigma^+$) polarized 1.5 ps pulses generated by a mode-locked
Ti-sapphire laser with a repetition rate of 80 MHz. The PL
intensities co-polarized $(I^+)$ and counter-polarized $(I^-)$
with the excitation laser are recorded in the backward direction
using a S1 photocathode Hamamatsu Streak Camera with an overall
time-resolution of 8 ps. We measure the PL circular polarization
degree $\rho = (I^+ - I^-)/(I^+ + I^-)$ proportional to
free-electron spin polarization $P_e= (n_+ -n_-)/(n_+ + n_-)$
where $n_+$ and $n_-$ are the densities of spin-up and spin-down
free electrons, and $n=n_+ + n_-$ is the free-electron total
density. Simultaneously, we measure the free-electron (total) spin
density $S_z=(n_+ - n_-)/2$, which for the bimolecular
recombination process writes: $S_z \propto (I^+ - I^-)/\sqrt{I}$,
where $I=I^+ + I^-$ is the total PL intensity.

\begin{figure}
  \centering
    \includegraphics[width=\linewidth]{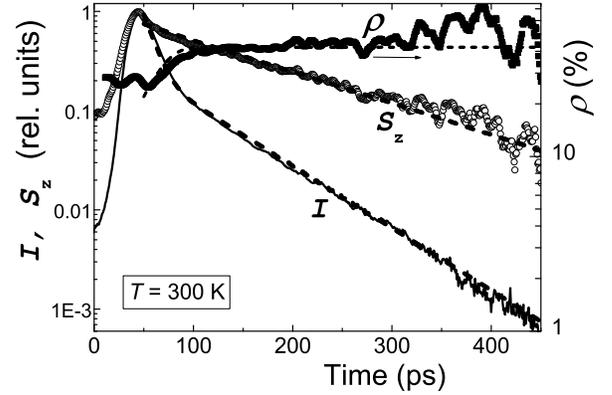}
  \caption{Time-resolved PL intensity (solid line),
  circular polarization (points)
  and total free-electron spin (circles) measured in
  GaAs$_{0.979}$N$_{0.021}$ alloy at room temperature. The
  laser excitation energy is \emph{h}$\nu_{exc}$ = 1.39 eV which corresponds
  to the photogeneration of carriers below GaAs barrier. The pump average power is $W= 130$ mW.
  Detection energy corresponds to the dominant conduction band $-$ heavy-hole
  recombination~\cite{Egorov}.
    Dashed curves are calculated in framework of the applied model (see text).}\label{f_scheme}
\end{figure}

Figure 1 shows the experimental kinetics $I(t)$ (solid line),
$\rho(t)$ (points) and $S_z(t)$ (circles) measured in
GaAs$_{0.979}$N$_{0.021}$ alloy at room temperature~\cite{note 1}.
One can see that $I$~and $S_z$ first rise for $\sim 45$ ps, which
results from relaxation and thermalization of photogenerated
carriers. During this time, $\rho$ varies slightly being close to
25\%. Then the PL decays in two steps: the first one is fast and
the second one is slower with the characteristic decay times of
$\approx15$ and $\approx70$ ps, respectively. Simultaneously with
the fast initial PL decay, the PL polarization monotonously rises
up to $\approx43$\% and afterwards keeps the value at least 2 ns
as we estimated considering noise fluctuations. Such a long
conservation of the high PL polarization is quite surprising since
at room temperature spin relaxation time of free electrons in
undoped bulk semiconductors of a zinc-bland structure is of the
order of magnitude of 100 ps~\cite{OO, Song}, i.e. 10 times
shorter than we observed. However, the total electron spin $S_z$
decays fast with a single characteristic time of about 150 ps.

\section{Theoretical model and discussion}
\begin{figure}[b]\centering
    \includegraphics[width=.35\textwidth]{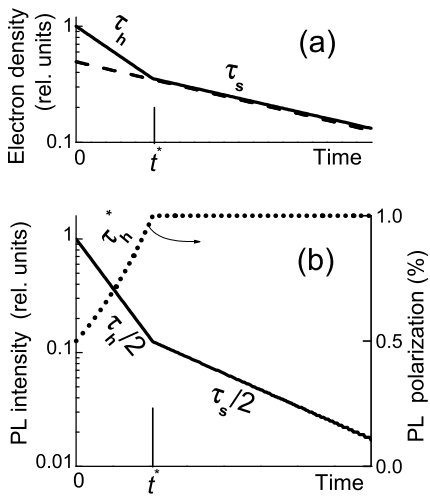}
        \caption{
Time evolution of (a) the total (solid line) and difference
(dashed) electron densities and (b) the PL intensity (solid) and
circular polarization (dotted), in schematic representation.
}\label{f_valleys}
\end{figure}
 Let us give first a qualitative explanation of the effects observed.
Figure 2 describes schematically two successive regimes of the
spin-dependent electron dynamics responsible for the formation of
high electron spin polarization and its persistence under the
circularly polarized interband pulsed photoexitation. We assume
that (i) the density $n_0$ of photoinjected electrons exceeds the
concentration $N_c$ of deep centers containing at equilibrium one
electron per center, (ii) the deep center can only capture a
photogenerated electron from the conduction band with a spin
antiparallel to the spin of the already present electron, forming
a singlet two-electron bound state~\cite{Weisbuch}, (iii) the
interband electron-hole recombination contributes negligibly to
the balance regulating the electron and hole densities and (iv)
photoinjected holes are unpolarized because of their fast spin
relaxation~\cite{OO}.

The first regime starts when all the centers acquire the second
electron, the fast electron capture is blocked and the further
capture is controlled by the photohole nonradiative recombination
with the bound electrons described by the hole lifetime $\tau_h =
(R_h N_c)^{-1}$, where $R_h$ is a constant. Thus, the electron and
hole densities decay as $n = (n_0 - N_c) \exp{(- t/\tau_h)}$, $p =
n_0 \exp{(- t/\tau_h)}$ (solid line in Fig. 2a). For the PL
intensity one has $I \propto \exp{(- 2t/\tau_h)}$. This regime
ends at the moment $t_{1}^*$ when the spin minority photoelectrons
disappear from the conduction band: $n_-(t_{1}^*) \approx0$. This
time can be estimated by $t_1^* \approx \tau_h \ln{(1/|P_i|)}$,
where $P_i$ is the initial degree of electron spin
polarization~\cite{note 3}. Within the time $t_{1}^*$ the minority
density exponentially decreases to zero while the density of the
majority electrons tends to a finite value $n_0 |P_i|$ which
results in a monotonously increasing free-electron spin
polarization up to $\approx 100\%$, see the dotted curve in
Fig.~2b.

The second regime is characterized by the almost unoccupied
minority spin conduction subband and the spin-polarized bound
electrons. The kinetics is governed by the free-electron spin
relaxation time $\tau_s$: the recombination rate for the free
electrons and holes is given by $n_+(t)/\tau_s$. This can be
understood taking into account the following balance of rates in
the second regime. The decrease of the electron majority density
due to the spin-flip processes is determined by the rate
$n_+(t)/(2\tau_s)$. The reverse-spin electrons are immediately
captured by deep centers which next capture unpolarized holes to
generate unpolarized paramagnetic centers. The majority free
electrons are captured at the generated paramagnetic centers with
the rate $n_+(t)/(2 \tau_s)$. This leads to the total decay rate
$n_+(t)/\tau_s$ for free photoelectrons, so that $n(t) \approx
n_+(t_{1}^*) \exp{(- t/\tau_s)}$, $I(t) \propto \exp{(-
2t/\tau_s)}$ while their spin polarization keeps on at the level
$\approx 100\%$. The densities of one-electron (paramagnetic) and
two-electron (nonmagnetic) centers, $N_1$ and $N_2$ ($N_2\equiv
N_c - N_1$), respectively, are constant. Particularly, $N_2
\approx (R_h \tau_s)^{-1}$ because the hole lifetime equals
$\tau_s$. The second regime ends at the moment $t_1^* + t_2^{*}$
when the decreasing free-electron density becomes smaller than
$N_2$ and the further kinetics is determined by the hole
recombination processes. The time $t_2^{*}$ is approximately given
by $\tau_s \ln{(n_0 |P_i| R_h \tau_s)}$ which can remarkably
exceed $\tau_s$. It is worth to stress that in contrast to $t_1^*$
the time $t_2^{*}$ is a function of the pump intensity and can be
varied in a wide range.

In contrast to the two-stage decrease of the free-electron density
$n(t)$ with two characteristic times $\tau_h$ and $\tau_s$ (solid
line in Fig.~2a), the spin-dependent difference $n_+(t)- n_-(t)$
decays with a single characteristic time $\tau_s$ (dashed line in
Fig.~2a). Therefore, the total spin $S_z = (n_+ - n_-)/2$ of free
electrons is controlled by the short time, $\tau_s$, of their spin
relaxation: $S_z(t)=S_z(0)\exp{(-t/\tau_s)}$.

The experimental data in Fig.~1 are obtained at the pump average
power of $W$ = 130 mW. At this power, as our estimations show (see
below), the ratio $n_0/N_c \sim 10$. In addition, the PL
exponential decay at long delay also indicates bimolecular
mechanism of recombination. This means that $n \approx p$ and  $n>
N_2$ up to 425 ps. Thus, the suggested model can be applied for
description of the experiment. Since $n \approx p$, the GaAsN
layer under study represents an intrinsic semiconductor where the
lifetimes of photogenerated electrons and holes coincide,
$\tau_e\approx \tau_h$. At the first stage, the PL decay
time-constant, $\tau_1$, is equal to $\tau_h/2 = (15.0\pm0.5)$ ps,
therefore, $\tau_e\approx \tau_h = (30\pm1)$ ps. At the second
stage, where $\tau_e\approx \tau_s$, the PL decay constant equals
$\tau_2 = \tau_s/2 = (72\pm2)$ ps, which yields $\tau_e \approx
\tau_h = (144\pm4)$ ps.

\begin{figure}[b]\centering
    \includegraphics[width=0.8\linewidth]{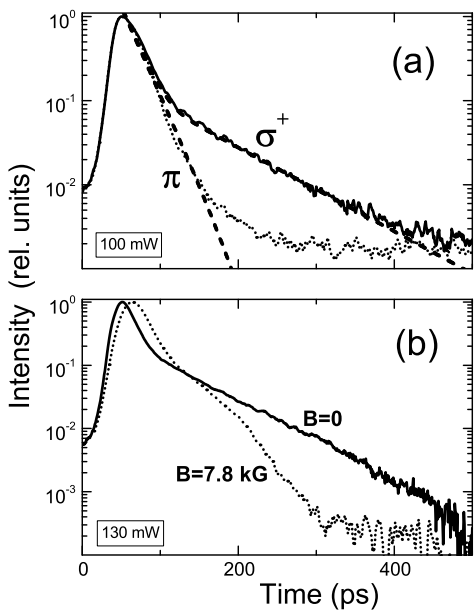}
  \caption{PL decay in GaAs$_{0.979}$N$_{0.021}$ alloy
  under $\sigma^+$- and $\pi$-polarized excitation (a) and
  under $\sigma^+$ excitation in the absence and presence of a
  transverse magnetic field (b). Solid and dotted curves are
  experimental transients,
  dashed curves are the result of calculation. \emph{W}(mW):
  100 (a), 130 (b).}\label{f_vo}
\end{figure}

The PL polarization is not sensitive to free-electron spin
relaxation, and, therefore, it cannot be used for measuring the
time $\tau_s$. However, the total free-electron spin $S_z$ decays
exponentially with the time constant $\tau_s$. The time was found
from the experimental dependency $S_z(t)$ shown in Fig.~1 to be
$\tau_s = (150\pm15)$ ps. Within the experimental error, this
value is two times longer than $\tau_2 =(72\pm2)$ ps which
coincides completely with the model predicting $\tau_s = 2\tau_2$.
Thus, the measurement of slow PL decay time at spin-dependent
recombination allows us to find the free-electron spin relaxation
time. Note, that the method is more precise as compared to the
measurement of the total spin decay where the ratio of $(I^+ -
I^-)/\sqrt{I}$ is found.

The suggested model is based on the dynamical polarization of deep
centers, which causes a decrease in the recombination rate of the
majority spin electrons at the second regime. If the deep-center
polarization is destroyed then the free-electron recombination
rate increases up to the one at the fast regime. The disappearance
of the slow range and the PL decay with the single short time
$\tau_1\sim16$ ps are observed under the change of the exciting
light polarization from the circular ($\sigma^+$) to linear
($\pi$) (Fig. 3a). Switching on the perpendicular magnetic field
of 7.8 kG under $\sigma^+$-pumping leads to the same effect (Fig.
3b). It again evidences the relevance of the suggested model.

To describe the experimental results quantitatively, we use the
set of equations~\cite{JETPLett} for the rates of the free- and
bound-electron total spins, respectively, ${\bm S}$ and ${\bm
S}_c$, transformed to the form:
\begin{eqnarray} \label{7II}
&& \frac{dn}{dt} + \frac{R_e}{2}\ (n N_1 - 4 {\bm S} {\bm S_c}) =
G\:, \nonumber\\
&& \frac{dp}{dt} + R_h (N_c - N_1) p = G\:, \nonumber\\
&&p = N_c - N_1 + n \:, \nonumber\\
&& \frac{d{\bm S}}{dt} + \frac{R_e}{2}\ ({\bm S} N_1 - {\bm S_c}
n) + \frac{{\bm S}}{\tau_s} + {\bm S} \times {\bm
\omega} = \frac{P_i}{2}\ G {\bm o}_z\:, \nonumber\\
&& \frac{d {\bm S_c}}{dt} + \frac{R_e}{2}\ ({\bm S_c} n - {\bm S}
N_1) + \frac{{\bm S_c}}{\tau_{sc}} + {\bm S_c} \times {\bm
\Omega}= 0\:.\nonumber
\end{eqnarray}
Here $G = G_+ + G_-$, $G_{\pm}$ are the photogeneration rates of
electrons with the spin $\pm 1/2$, $P_i = (G_+ - G_-)/G$,
$\tau_{sc}$ is spin relaxation time of bound electrons, ${\bm
o}_z$ the unit vector directed along the growth axis $z$
coinciding with the exciting beam, ${\bm \omega}, {\bm \Omega}$
are the Larmor frequencies defined by $\hbar {\bm \omega} = g
\mu_B {\bm B}$, $\hbar {\bm \Omega} = g_c \mu_B {\bm B}$, $g$ and
$g_c$ are the $g$-factors of free and bound electrons, ${\bm B}$
is the magnetic field. Note that $N_1 = N_+ + N_-\:,\:S_{c,z}=
(N_+ - N_-)/2$, where $N_{\pm}$ are the densities of spin-up and
spin-down paramagnetic centers, and we assume ${\bm B} \perp z$.
The sign of experimentally observed positive PL polarization in
Fig.~1 corresponds to the recombination of conduction electrons
with heavy holes~\cite{Egorov}. For the recombination, $\rho= P'
P_e = 2P' S_z /n$~\cite{OO}, where $P'$ is a depolarization
factor.

Our model considers kinetics when the PL decay starts. We assume
that the decay begins at the time delay $\Delta t = 55$ ps which
is used as a fitting parameter. Calculated dependencies of the PL
intensity $I(t)$, degree of circular polarization $\rho(t)$ and
spin $S_z(t)$ shown by dashed curves in Fig.~1 and Fig.~3a are
obtained for the following parameters: $n_0/N_c=10$ and 7.7 for
Fig. 1 and Fig. 3a, respectively, $R_e/R_h=4$, $\tau_{s}=140$ ps,
$\tau_{sc}$~=~1500 ps, $P_i=50$ and $0\%$ at $\sigma^+$ and $\pi$
excitations. One can see, that the calculated curves fit well the
experimental ones. Note that the average value of $\rho$ measured
within the plateau in Fig. 1 equals 43\%. It is two times smaller
than the value, 95\%, calculated assuming recombination of
conduction electrons with heavy holes. We attribute the difference
in the $\rho$ value to admixture of the conduction
electron$-$light-hole recombination having the opposite sign of
$\rho$~\cite{Egorov}. A coincidence of the calculated (dashed)
curve $\rho(\emph{t})$ with the experimental one in Fig.~1 is
obtained using the fitting coefficient of $P' = 0.452$.

The calculations (not presented here) show that the normalized
relations $\rho(\emph{t})$ and $I(\emph{t})$ depend weakly on
variation of the parameters $n_0/N_c$ and $R_e/R_h$. Specifically,
they virtually coincide at $n_0/N_c >20$ and $R_e/R_h>6$. At the
same time, we found that calculated for a continuous-wave (CW)
excitation dependencies $\emph{I}$ and $\rho$ on both the
excitation intensity and transverse magnetic field (Hanle effect)
are very sensitive to these parameters. Fitting of experimental
curves found in Ref.~[\cite{JETPLett}] at CW excitation of the
same GaAs$_{0.979}$N$_{0.021}$ sample allowed us to estimate the
value of $N_c \sim{4}\cdot10^{16}~ $cm$^{-3}$. This provided in
turn to find the relation of $n_0/N_c\sim10$ and $R_e/R_h\sim4$ at
the average pulse pump power $\emph{W}=130$ mW.

The value $\tau_{sc}=1500$ ps used in this paper is in agreement
with the 600 ps estimate from below for $\tau_{sc}$ found in
Ref.~[\cite{JETPLett}] from the analysis of the Hanle effect. It
should, however, be mentioned that, in comparison to the
continuous-wave measurements of Ref.~[\cite{JETPLett}], the
time-resolved data are less sensitive to $\tau_{sc}$.
Particularly, the ``plateau'' of the curve $\rho(\emph{t})$ lasts
out up to 500 ps even for $\tau_{sc}$ decreased to 30 ps which
clearly demonstrates dynamical character of the observed
spin-dependent effects.

\begin{figure}
  \centering
    \includegraphics[width=1.0\linewidth]{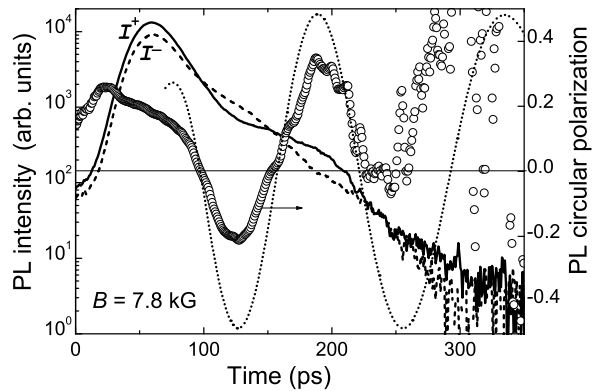}
\caption{Room-temperature electron spin beats in the PL intensity
$\sigma^+$ (solid) and $\sigma^-$ (dashed) components and the
degree of PL circular polarization (circles). $W= 130$ mW. Dotted
curve is the calculated one.} \label{f_so_k}
\end{figure}

Fig.~4 shows the beats in the $I^+$ (solid) and $I^-$ (dashed) PL
components and in the polarization degree $\rho$ (circles)
recorded in GaAs$_{0.079}$N$_{0.021}$ alloy in the presence of the
transverse magnetic field $\emph{B} = 7.8$~kG. These beats result
from the Larmor precession of electron spins. They have a complex
shape since $g$-factors of free and bound electrons, $g$ and
$g_c$, differ both in their value and sign. However, the exact
values of $g$ and $g_c$ in GaAs$_{0.079}$N$_{0.021}$ are unknown.
Since in our measurements~\cite{JETPLett} the sign of $g_c$ was
found to be positive, we believe that the $g$-factor of bound
electrons is close to the $g$-factor of electrons in vacuum, i.e.,
$g_c \approx 2$. As for $g$-factor of free electrons, a paper
devoted to measuring~~$g$ in GaInAsN dilute nitrides is known
only~\cite{Skierb}. It was found in this paper that $g$ is
negative. We used $g$-factor $g$ = $-$0.9 in our calculations
since at this value the calculated beat period fits the
experimental one. It is seen in Fig.~3b that the peak $I(t)$ is
delayed by 14 ps in the magnetic field 7.8 kG as compared to the
case of \emph{B}=0. In accordance to this, we assume that, in the
presence of magnetic field, $\Delta t = 69$ ps. Dotted line in
Fig. 4 shows the beats of $\rho$ calculated with $g$ = $-$0.9,
$g_c = 2$ and $\Delta t = 69$ ps. One can see that our model
describes qualitatively the experimental beats for $t > 70$ ps.

We have also observed strong PL polarization and its stability at
room temperature in an $n$-doped ($N_{\rm Si} \sim 10^{17}~
$cm$^{-3}$) GaAsN layer and in an undoped compressively strained
quantum well InGaAsN/GaAs with small nitrogen content of the order
of 1\%~\cite{Lombez}. This indicates the general character of the
observed effects.

To conclude, we have measured the giant polarization ($\sim45\%$)
of free electrons and its persistence ($> 2$ ns) in GaAsN alloys
at room temperature. We have developed the nonlinear theory of
spin dynamics in a coupled system of free and localized carriers
controlled by spin-dependent recombination. The theory shows that
the increase of both free and localized electron polarizations up
to their limiting values is due to dynamical polarization of deep
paramagnetic centers. When the latter is effective, the mean spin
of free electrons is independent of their spin relaxation while
the total free-electron spin decays in the conduction band with
the short spin relaxation time of $\sim100$ ps. Also, electron
spin quantum beats are observed in transverse magnetic field at
room temperature and described theoretically.

The work is partially supported by the Russian Foundation for
Basic Research and Programs of the Russian Academy of Sciences.


\begin{thebibliography}{99}

\bibitem{Weisbuch} C.~Weisbuch and G.~Lampel, Solid State Commun.
{\bf 14}, 141 (1974).

\bibitem{Miller} R.C.~Miller, W.T.~Tsang, and W.A.~Nordland, Jr., Phys. Rev. B {\bf 21}, 1569 (1980).

\bibitem{Paget} D.~Paget, Phys. Rev. B {\bf 30}, 931 (1984).

\bibitem{JETPLett} V.K.~Kalevich, E.L.~Ivchenko, M.M.~Afanasiev, A.Yu.~Shiryaev,
A.Yu.~Egorov, V.M.~Ustinov, B.~Pal, and
Y.~Masumoto, JETP Lett. {\bf 82}, 455 (2005).

\bibitem{Egorov} A.Yu.~Egorov, V.K.~Kalevich, M.M.~Afanasiev, A.Yu.~Shiryaev,
V.M.~Ustinov, M.~Ikezawa, and Y.~Masumoto, J.
Appl. Phys. {\bf 98}, 13539 (2005).

\bibitem{note 1} Experimental kinetics obtained in the samples with
the $x$ = 2.7 and 3.4\% are qualitatevely the same.

\bibitem{OO} {\it Optical orientation}, eds. F.~Meier and B.~Zakharchenya
(North-Holland, Amsterdam, 1984).

\bibitem{Song} Pil Hun Song and K.W.~Kim, Phys. Rev. B {\bf 66}, 35207 (2002).

\bibitem{note 3} Taking into account free-electron spin relaxation,
$t_1^* =\tau_h^*\ln{(1/|P_i|)}$, where $(\tau_h^*)^{-1} =
(\tau_h)^{-1}-(\tau_s)^{-1}$. In our sample during the first
regime, $\tau_s\approx5\tau_h$, therefore $\tau_h^*\approx\tau_h$
and $t_1^* \approx\tau_h \ln{(1/|P_i|)}$.

\bibitem{Skierb} C.~Skierbiszewski, P.~Pfeffer, and J.~Lusakowski, Phys. Rev. B {\bf 71}, 205203 (2005).

\bibitem{Lombez} L.~Lombez, P.-F.~Braun, H.~Carrere, B.~Urbaszek, P.~Renucci,
T.~Amand, X.~Marie, J.C.~Harmand, and V.K.~Kalevich, Appl. Phys.
Lett. {\bf 87}, 252115 (2005).

\end{thebibliography}
\end{document}